\documentclass[twocolumn,aps,prl,showpacs,superscriptaddress]{revtex4}
\usepackage{graphics,bm}
\usepackage{amssymb}
\usepackage{epsfig}
\usepackage{epsf}
\usepackage{color}

\def\be{\begin{equation}} \def\ee{\end{equation}}
\def\beq{\begin{eqnarray}} \def\eeq{\end{eqnarray}}

\def\nn{\nonumber}

\begin{document}

\title{Tuning a Strain-Induced Orbital Selective Mott Transition in Epitaxial VO$_2$}
\author{Shantanu Mukherjee*}
\affiliation{Department of Physics, Applied Physics and Astronomy, Binghamton University, Binghamton, New York 13902, USA}

\author{N. F. Quackenbush*} \affiliation{Department of Physics, Applied Physics and Astronomy, Binghamton University, Binghamton, New York 13902, USA}

\author{H. Paik}\affiliation{Department of Materials Science and Engineering, Cornell University, Ithaca, New York 14853-1501, USA}

\author{C. Schlueter}\affiliation{Diamond Light Source Ltd., Diamond House, Harwell Science and Innovation Campus, Didcot, Oxfordshire OX11 0DE, UK}

\author{T.-L. Lee}\affiliation{Diamond Light Source Ltd., Diamond House, Harwell Science and Innovation Campus, Didcot, Oxfordshire OX11 0DE, UK}

\author{D. G. Schlom}\affiliation{Department of Materials Science and Engineering, Cornell University, Ithaca, New York 14853-1501, USA}\affiliation{Kavli Institute at Cornell for Nanoscale Science, Ithaca, New York 14853, USA}

\author{L. F. J. Piper}\email{lpiper@binghamton.edu} \affiliation{Department of Physics, Applied Physics and Astronomy, Binghamton University, Binghamton, New York 13902, USA} \affiliation{Materials Science {\&} Engineering, Binghamton University, Binghamton, New York 13902, USA}

\author{Wei-Cheng Lee}
\email{wlee@binghamton.edu}
\affiliation{Department of Physics, Applied Physics and Astronomy, Binghamton University, Binghamton, New York 13902, USA}

\date{\today}

\begin{abstract}
We present evidence of strain-induced modulation of electron correlation effects and increased orbital anisotropy in the rutile phase of epitaxial VO$_2$/TiO$_2$ films from hard x-ray photoelectron spectroscopy and soft V L-edge x-ray absorption spectroscopy, respectively. By using the U(1) slave spin formalism, we further argue that the observed anisotropic correlation effects can be understood by a model of orbital selective Mott transition at a filling that is non-integer, but close to the half-filling. Because the overlaps of wave functions between $d$ orbitals are modified by the strain, orbitally-dependent renormalizations of the bandwidths and the crystal fields occur with the application of strain. These renormalizations generally result in different occupation numbers in different orbitals. We find that if the system has a non-integer filling number near the half-filling such as for VO$_2$, certain orbitals could reach an occupation number closer to half-filling under the strain, resulting in a strong reduction in the quasiparticle weight $Z_{\alpha}$ of that orbital. Moreover, an orbital selective Mott transition, defined as the case with $Z_{\alpha} = 0$ in some, but not all orbitals, 
could be accessed by epitaxial strain-engineering of correlated electron systems. 

*{\it These authors contribute equally.}
\end{abstract}


\maketitle

{\it Introduction} --
Mott insulators are characterized by the ratio of $U/W$, where $U$ refers to the correlation strength and $W$ is the full bandwidth. A Mott state is usually predicted from theoretical models to exist in materials at half filling and when $U/W\sim 1$. 
However in multi-orbital systems an incipient Mott picture has been proposed with the possibility of an ``orbital selective Mott transition" (OSMT). The OSMT scenario was first proposed to understand the physical properties of Ca$_{2-x}$Sr$_x$RuO$_4$ by Anisimov\cite{anisimov2002}, stimulating a lot of research efforts on this subject.\cite{liebsch2003,liebsch2004,fang2004,koga2004, demedici2005} More recently, in the iron based superconductors, OSMT has been suggested to account for the insulating properties driven by the iron vacancy in K$_{1-x}$Fe$_{2-y}$Se$_2$. \cite{demedici2009,demedici2011,yu2011,yu2012,yi2013,julian2014,demedici2014,giovannetti2015} In general, the physics of OSMT has relevance to a variety of 
multi-band materials that are close to a Mott transition, 
and a better understanding of this phase both theoretically and experimentally would be crucial for advancing our understanding of Mott physics. 

Vanadium dioxide (VO$_2$) is one of the early prototypes for strongly correlated systems near half filling.\cite{imada1998,haverkort2005} VO$_2$ exhibits a metal to insulator transition (MIT) with a concomitant formation of V-V dimers that is generally considered to be driven cooperatively by both Mott and Peierls physics.\cite{morin1959,goodenough1971,eyert2002,haverkort2005,koethe2006} For example, it has been pointed out that the massive orbital switching that occurs upon entering the insulating phase can only be achieved if the system is already close to a Mott insulating regime.\cite{haverkort2005} A significant theoretical effort has been made to understand this MIT using local density approximation (LDA), LDA + U, LDA + DMFT, etc.
\cite{eyert2002,haverkort2005, biermann2005, weber2012}  Specifically, a Peierls-assisted OSMT mechanism has been proposed to understand the MIT between metallic $R$ and insulating $M1$ phases in bulk VO$_2$.\cite{weber2012} Furthermore, recent progress in ultra thin epitaxial VO$_2$ growth using TiO$_2$ substrates has provided surmounting evidence that electron correlation effects may become enhanced by the large strains, thus pushing the system further into the Mott regime without introducing any dopants.\cite{paik2015, quackenbush2015,quackenbush2016,laverock2012}. These results suggest that VO$_2$ is a candidate for exploring the OSMT with epitaxial strain.

In this Letter we present a generalized theory of OSMT to describe how strain can result in quasiparticle weight variation and preferential orbital switching. We perform theoretical calculations to show that an OSMT can be tuned within the metallic phase of VO$_2$ (R-phase) by applying a strain that enhances the electron correlations. We find that the application of a uniaxial strain tends to make the occupation number on each orbital unequal. As a result, even if the system is slightly away from half filling, the presence of a moderate anisotropic strain along particular directions could render the occupation number in certain orbitals closer to half-filling, leading to an OSMT. Further, by performing hard x-ray photoelectron spectroscopy  (HAXPES) and V L-edge x-ray absorption spectroscopy (XAS) on 10 nm epitaxial VO$_2$ films on TiO$_2$(001), (100) and (110), referred to as VO$_2$(001), VO$_2$(100), and VO$_2$(110) (see Supplementary), we show that strain induced electronic effects indeed exist in the metallic phase of VO$_2$, in an agreement with our theoretical predictions. 

{\it Model and Formalism} --
We have employed the $U(1)$ slave spin formalism\cite{yu2012} on a two orbital model to investigate the effect of strain on the quasiparticle weight near the transition to the featureless Mott phase. The slave spin formalism\cite{demedici2005} has been shown to reproduce the Mott transition at the mean-field level in good agreement with DMFT results, and the $U(1)$ version can even obtain the correct non-interacing limit at the mean-field level.\cite{yu2011} We start from a generic two orbital model containing a tight binding Hamiltonian and a multi-orbital Hubbard interaction term representing 
onsite electron correlations (see Supplementary Materials). Below we briefly summarize the mechanism for studying OSMT in a $U(1)$ slave spin formalism\cite{yu2012} applied to our model Hamiltonian.

Representing the charge degree of freedom by a quantum spin 1/2, the electron creation operator $d^\dagger_{i\alpha\sigma}$ can be written as
$
d^\dagger_{i\alpha\sigma} = S^+_{i\alpha\sigma} f^\dagger_{i\alpha\sigma},
$
where $d^\dagger_{i\alpha\sigma}$ creates an electron on the orbital $\alpha$ with spin $\sigma$ at site $i$, $S^+_{i\alpha\sigma}$ is the spin raising operator for the slave spin describing a charge on the orbital $\alpha$ with physical spin $\sigma$ on the site i, and 
$f^\dagger_{i\alpha\sigma}$ is the fermionic spinon associated with the physical spin. 
After solving the mean-field equations subject to the constraints to eliminate the unphysical Hilbert spaces due to the introduction of the slave spins, the quasiparticle weight on orbital $\alpha$ can be obtained by $Z_{\alpha\sigma}\propto \vert\langle S^+_{\alpha\sigma}\rangle\vert^2$.\cite{yu2012} Finally, the correlated metallic phase (Mott insulating phase) corresponds to $Z_{\alpha\sigma}\neq 0$ ($Z_{\alpha\sigma}= 0$) for every orbital, and the OSMT is given by $Z_{\alpha\sigma}=0$ for some orbitals. Note that the OSMT is not a complete insulating state since the quasiparticle weight in certain bands is still finite. We perform the calculations at zero temperature and limit our attention only on the featureless Mott insulating phases. As a result, we have $\langle n_{\alpha\sigma}\rangle = \frac{n_\alpha}{2}$ and $Z_{\alpha\sigma} = Z_{\alpha}$. In a two orbital model, the half-filling is corresponding to $n = n_1 + n_2 = 2$, and we focus on the case with $1<n<2$.

{\it Results} --
Although the generic tight binding model can describe any two orbital system, we begin by choosing a $d_{xz}$ and $d_{yz}$ two orbital system for the purpose of demonstration 
(see Supplementary Materials). Now assume that the strain is applied to elongate the system along $\hat{x}$ direction. This strain significantly reduces the wave function overlaps associated with the $d_{xz}$ but has much less 
effect on the wave function overlaps associated with the $d_{yz}$. As a result, the bandwidth of $d_{xz}$ orbital should reduce, and we may introduce an effective parameter $d$ such that 
$
t^{11,strain}_{\vec{k}}\equiv (1-d) t^{11}_{\vec{k}}
$
Here $t^{11}$ is the hopping matrix element associated with the $d_{xz}$ orbital and $(1-d)$ reflects the bandwidth reduction in the $d_{xz}$ orbital.
As a test, we have studied the case with $C_4$ symmetry and $d = 0$, which is summarized in Supplementary Materials.
\begin{figure}
	\includegraphics[width=3.4in]{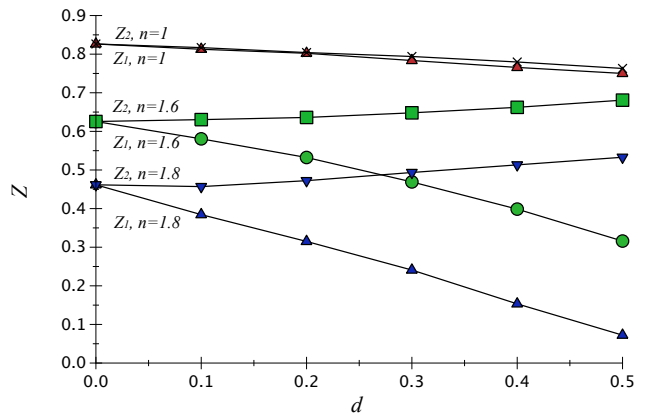}
	\caption{\label{fig:strain} The orbitally-dependent quasiparticle weight in the case under the strain with $U=4.5$ for different filling $n$.
		At the filling away from the half-filling ($n=1$), the quasiparticle weights remain almost the same 
		even at large $d$. In contrast, at the filling near the half-filling ($n=1.8$ and $n=1.9$), the quasiparticle weight in $d_{xz}$ ($Z_1$) reduces significantly as $d$ increases.}
\end{figure}

Now we turn on the strain by setting $d \neq 0$. There are two effects of $d$ on the $d_{xz}$ orbital. First, since the its bandwidth $W_1$ is reduced, the ratio of the interaction 
to the bandwidth $U/W_1$ increases, leading to stronger correlation effect on $d_{xz}$ orbital. Second, if $n < 2$, the reduction of the bandwidth $W_1$ tends to attract more electrons 
to occupy $d_{xz}$ orbital. Therefore we can expect that under the strain, $n_1 = n/2 + x$ and $n_2 = n/2 - x$. This indicates that $d_{xz}$ orbital is pushed even closer to the 
half-filling and consequently much more correlated. As a result, both effects favor driving $d_{xz}$ more correlated, and an OSMT could be obtained 
if $n$ is near the half-filling.

The orbitally-dependent quasiparticle weight $Z_\alpha$ as a function of strain with fixed $U$ at different $n$ is presented in Fig. \ref{fig:strain}.
For $n=1$, the strain does not affect $Z_{\alpha}$ too much even at quite large $d$. 
As $n$ is closer to 2, the difference between $Z_1$ and $Z_2$ becomes more prominent, consistent with our discussion above.
It should be noted that this result is quite general for a multi-orbital system, and an intriguing point is that one could engineer the OSMT by the strain effect 
even in a system with non-integer fillings. In the next section, we study the MIT in VO$_2$ under high strain using the same formalism developed above.

{\it Application to epitaxially strained VO$_2$} --
Although the origin of the MIT in VO$_2$ is a complicated issue, the dominant bands near the Fermi surface are well-known. The orbital compositions of the three 
dominant bands near the Fermi surface can be expressed as $\vert\sigma (d_\parallel)\rangle = \vert x^2-y^2\rangle$, and 
$\vert\pi_{\pm}\rangle = \frac{1}{\sqrt{2}}\left(\vert xz\rangle\pm\vert yz\rangle\right)$.
While the leading hopping parameter of $\sigma$ band mostly comes from the strong direct $\sigma$ bonding between nearest $V-V$ pairs, the leading hopping parameters of $\pi_\pm$ bands come from a combination of the direct $\pi$ bonding between nearest $V-V$ pairs and the second order hoppings via the $V-O$ bonds between $3d$ and $2p$ orbitals.\cite{tanaka2004,haverkort2005} 
Tanaka estimated\cite{tanaka2004} $\vert t^{\sigma\sigma}\vert \sim \vert t^{\pi_-\pi_-}\vert >> \vert t^{\pi_+\pi_+}\vert$, which is in close agreement with LDA calculations.\cite{haverkort2005}
Moreover, due to the crystal field effect coming from the oxygen atoms, $\pi_\pm$ bands are pushed to the higher energy, leading to a lower onsite energy in the $\sigma$ band. As a result, this system can be reasonably approximated as a two-band system with comparable bandwidths in the metallic $R$ phase. A strain on VO$_2$ due to the (110) or (100)-oriented TiO$_2$ substrate leads to two important effects that support the OSMT scenario. 
Firstly, because the $c$-axis lattice constant of the bulk TiO$_2$ is longer than that of bulk VO$_2$ by $3.62\%$\cite{muraoka2002}, the $c$-axis is stretched in 
both VO$_2$(110) and VO$_2$(100) (note that the $\hat{x}$-direction in the tetragonal unit cell is along the $c$-axis\cite{eyert2002,lazarovits2010}). 
It is found in LDA calculations that a longer $c$-axis primarily leads to a reduction of the $\sigma$ orbital hoppings $t^{\sigma\sigma}$ and suppression of the corresponding band width $W_{\sigma}$.\cite{lazarovits2010} Consequently a larger U/$W_{\sigma}$ would enhance the electronic correlations in $\sigma$ band. 
Secondly, due to the elongation of $c$-axis, the average $V-O$ bond length would shorten, which enhances the $d-p$ hybridizations, increases the on-site energy of $\pi$ orbital, and leads to a larger 
energy difference between the $\sigma$ and $\pi$ orbitals.
For simplicity, we model the change in the onsite energy as an effective lowering of $\sigma$ orbital on-site energy. 
The effects of the strain can then be taken into account by
\beq
t^{\sigma\sigma,strain}_{\vec{k}}&\equiv& (1-d) t^{\sigma\sigma}_{\vec{k}}\nn\\
\Delta_\sigma^{strain} &=& - (1+Cd)\Delta_\sigma,
\label{strain-vo2}
\eeq
where $C$ is a positive constant. It can be seen easily that both effects from the strain strongly favor more occupation numbers in the $\sigma$ band, thus pushing it closer to half filling.
\begin{figure}
	\includegraphics[width=3.4in]{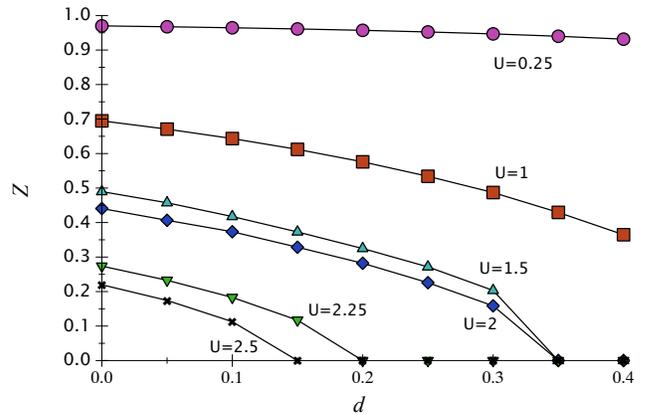}
	\caption{\label{fig:vo2} The quasiparticle weight in the $\sigma$ band of VO$_2$ under the strain as a function of $U$ at $n=1.5$.
		The hopping parameters for VO$_2$ are $t^{\sigma\sigma}_{i,i+\hat{x}} =t^{\sigma\sigma}_{i,i+\hat{y}} = -0.25$, $t^{\pi_-\pi_-}_{i,i+\hat{x}} 
		=t^{\pi_-\pi_-}_{i,i+\hat{y}} = 0.22$ . The onsite energy due to the crystal field splitting without the strain are
		$\Delta_\sigma = - 0.1$ and $\Delta_\pi = 0.05$. These values are taken from Ref. [\onlinecite{tanaka2004}], and the unit of energy is eV.
		The interband hoppings are small and neglected in Ref. [\onlinecite{tanaka2004}], and we set a representative value 
		$t^{\sigma\pi_-}_{i,i+\hat{x}+\hat{y}} = t^{\pi_-\sigma}_{i,i+\hat{x}+\hat{y}} = 0.01$.
		At large $U$, an OSMT could occur at small to moderate $d$.}
\end{figure} 

Moreover, because of the large hybridization between $3d$ orbitals of $V$ and $2p$ orbitals of $O$, the total occupation number in $3d$ orbitals is in fact in the range of 
$n = 1.2 - 2$ instead of 1 as one might expect from a direct counting of the valence charges.
Consequently, the OSMT due to the strain effect discussed in the last section is very likely to occur with a moderate $d$.
We adopt the values of $t^{\sigma\sigma}$, $t^{\pi_-\pi_-}$, and the onsite energies derived in Ref. [\onlinecite{tanaka2004}] and Eq. \ref{strain-vo2} to model the strain effect. 
The quasiparticle weight in the $\sigma$ band $Z_\sigma$ at $n = 1.5$ as a function of $U$ and $d$ is plotted in Fig. \ref{fig:vo2}.
For small $U$, $Z_\sigma$ remains close to 1 and does not depend on $d$. As $U$ increases, $Z_\sigma$ develops a significant dependence on $d$, and an orbital selective 
Mott transition occurs at a critical $d=d_c$ if $U$ is large enough. Clearly, $d_c$ decreases as $U$ increases, indicating that the OSMT is indeed driven by 
the strong Coulomb repulsive interaction. Furthermore, we notice that the behavior would be fundamentally different when the $c$-axis is compressed rather than elongated, as is the case for VO$_2$(001). This will shorten $\sigma$ bond distance along the $c$-direction and would reduce the energy difference between $\sigma$ and $\pi$ orbitals, thus keeping their occupations away from half filling in the R phase. As a result, the OSMT would be expected to occur in VO$_2$(100) and VO$_2$(110), but not in VO$_2$(001).

We emphasize that the increase of the occupation number in the $\sigma$ band is a direct consequence of the balance between kinetic and interaction energies. Because the bandwidth and the effective onsite energy of the $\sigma$ band are both lowered by the strain, it is energetically favorable to put more electrons in the $\sigma$ than in the $\pi_-$ bands to compromise the interaction energy.
While this trend in the occupation numbers can be captured by LDA-based approaches, the significant change in the quasiparticle weight as well as the OSMT found here is attributed to the advantage of the $U(1)$ slave spin formalism.
We predict that the occupation number in the $\sigma$ band could be substantially increased under the strain even in R phase without the dimerization of $V-V$ pairs, 
and the metallic phase in this region would be a strongly correlated metallic state with vanishing quasiparticle weight in the $\sigma$ band $Z_{\sigma}$. 

To investigate the predicted effect of strain on VO$_2$ thin films, we have performed HAXPES and polarization dependent XAS of the V L-edge on VO$_2$(001), VO$_2$(100), and VO$_2$(110) in the metallic R phase close to the MIT (see Supplementary). Figure \ref{fig:PES}a shows the HAXPES of the topmost valence states, while fig. \ref{fig:PES}b shows the corresponding V L XAS spectra collected from each VO$_2$ film. All presented spectra are collected at $\sim$20~$^\circ$C above T$_{MIT}$, i.e. in the metallic phase. The XAS at the V L-edge were collected with the polarization vector either parallel to the [001] (E$_\parallel$c$_R$) or [110] (E$_\perp$c$_R$) crystallographic axis. Included at the bottom of fig. \ref{fig:PES}b are the difference spectra (E$_\parallel$c$_R$~-~E$_\perp$c$_R$) representing the linear orbital dichroism. 

Both the valence band HAXPES and weak orbital dichroism of the VO$_2$(001) closely resemble that of bulk VO$_2$.\cite{koethe2006, haverkort2005, quackenbush2013} 
In contrast, both the VO$_2$(100) and (110) films show dramatic deviations from the bulk. 
The sharp Fermi edge at $E_F$, indicating Fermi liquid behavior, is weakened in the VO$_2$(100) case and further so for the (110) case resulting in a more smeared out intensity distribution near the Fermi energy. This behavior is equivalent to the vanishing $Z_\sigma$ in our model and is typically seen in Mott insulating materials such as in parent cuprate superconductors.\cite{shen1987,imada1998} Along with these changes at $E_F$ are concurrent increases in orbital anisotropy observed in the V L-edge spectra of the VO$_2$(100) and VO$_2$(110) films, indicating preferential filling of certain bands. This demonstrates that the strain, specifically when the $c$-axis is elongated, can indeed effect both the electron correlations and the orbital occupancy.

\begin{figure}[t]
	\includegraphics[width=3.2in]{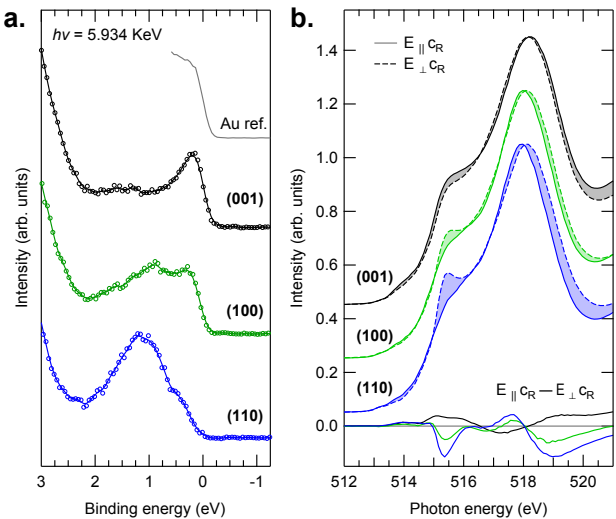}
	\caption{\label{fig:PES} a. HAXPES of the metallic phases of the three VO$_2$ strain orientations close to the Fermi level.  b. The corresponding polarization dependent V L-edge XAS for the films, along with the orbital dichroism for each case plotted underneath on a common intensity scale.}
\end{figure} 

The orbital dichroism observed here in the VO$_2$(100) and VO$_2$(110) films tends toward that observed for the insulating phase of unstrained VO$_2$, despite remaining in the metallic phase. \textcite{haverkort2005} observed a dramatic switching of the orbital occupancy going from the high temperature rutile to low temperature M1 phase, in which the $\sigma$ (often referred to as d$_\parallel$) orbital becomes preferentially filled. Our results indicate a similar modification of the orbital occupancy purely induced by strain within the distorted rutile structure. Remarkably, the increased preferential filling of the $\sigma$ orbital in this case does not require the formation of V--V dimers, supporting our OSMT picture.

Consistent with our OSMT prediction for VO$_2$, the weak orbital dichroism and the sharp Fermi edge for VO$_2$(001), when the $c$-axis is contracted, resemble that of the bulk. Furthermore, we observe that while both VO$_2$(110) and VO$_2$(100) have the same stretched $c$-axis, VO$_2$(110) exhibits larger orbital dichroism, as well as a weaker Fermi edge, indicating that 
VO$_2$(110) is more correlated than VO$_2$(100).
This difference can be attributed to the compression along the out-of-plane direction of the film due to the compensation of the biaxial strain, which is ignored in our theory. 
The compression in VO$_2$(110) is along rutile (110) direction (the $\hat{y}$-direction in the unit cell)\cite{eyert2002,lazarovits2010}, thus directly effecting the V-O bond lengths. 
This serves to increase the on-site energy of $\pi$ orbital due to the enhanced $d-p$ hybridization, which leads to a larger energy difference between the $\sigma$ and $\pi$ orbitals. 
However, in VO$_2$(100) the compression decreases the distance between $V$ atoms along the rutile $a$-axis (the $(\hat{y}-\hat{z})$ direction in the unit cell)\cite{eyert2002,lazarovits2010}, 
having a less direct effect on the V-O bond lengths. As a result, the strain effects distinguishing $\sigma$ and $\pi$ orbitals are expected to be stronger in VO$_2$(110). More rigorous theoretical efforts are required to obtain a quantitative description in this aspect.

{\it Conclusion} --
To summarize, we have studied the effect of strain on multiorbital systems. Because the spatial profile of the orbital wave function are anisotropic, the 
strain-induced bandwidth reduction and the onsite energy due to the crystal field effect become orbitally-dependent. 
Based on the $U(1)$ slave spin formalism on a generic two orbital Hubbard model, we have demonstrated that if the system has a total filling which is non-integer, but close 
to the half-filling, an orbital selective Mott transition could be engineered by the application of strain if the system has strong Coulomb repulsive interactions.
By applying this theory to study the highly strained VO$_2$ in combination with experimental results from spectroscopic studies, we have proposed that an orbital selective Mott state exists in the high temperature phase of coherently strained VO$_2$ with an elongated $c$-axis.
This state could appear before the M$_2$ phase emerges, which explains why the region of M$_2$ phase found in experiments is much narrower than the prediction of 
Landau-Ginzburg theory.\cite{quackenbush2016}.
Our results indicate that Mott physics is crucial to understand the properties of VO$_2$.


{\it Acknowledgement} --
We thank Dr. D. O. Scanlon for assisting with the experiments. S. M. and W.-C. L. acknowledge support from a start up fund from Binghamton University. L. F. J. P. and N. F. Q. acknowledge support from the National Science Foundation under DMR 1409912. The work of H.P. and D.G.S. was supported in part by the Center for Low Energy Systems Technology (LEAST), one of the six SRC STARnet Centers, sponsored by MARCO and DARPA. We thank Diamond Light Source for access to beamline I09 (SI12546) that contributed to the results presented here.

\newpage
\section{Supplementary Materials}
\subsection{Model}
A generic two orbital model can be written as
\beq
H&=& H_0 + H_U + H_{pair},\nn\\
H_0&=& \sum_{\alpha\beta\sigma}\sum_{ij} \left[-t^{\alpha\beta}_{ij} + \delta_{\alpha\beta}\delta_{ij}(\Delta_\alpha - \mu) \right]d^\dagger_{i\alpha\sigma} d_{j\beta\sigma}, \nn\\
H_U &=& U\sum_{i\alpha} n_{i\alpha\uparrow} n_{i\alpha\downarrow}
+ U'\sum_{i,\alpha < \beta,\sigma} n_{i\alpha\sigma}n_{i\beta-\sigma}\nn\\
&+& (U'-J) \sum_{i,\alpha < \beta,\sigma} n_{i\alpha\sigma}n_{i\beta\sigma},\nn\\
H_{pair} &=&- J\sum_i \big[d^\dagger_{i1\uparrow} d_{i1\downarrow} d^\dagger_{i2\downarrow}d_{i2\uparrow} + d^\dagger_{i1\uparrow}d^\dagger_{i1\downarrow} 
d_{i2\downarrow}d_{i2\uparrow}h.c.\big],\nn\\
\label{model}
\eeq
where $d^\dagger_{i\alpha\sigma}$ creates an electron on the orbital $\alpha$ with spin $\sigma$ at site $i$, $n_{i\alpha\sigma} = d^\dagger_{i\alpha\sigma}d_{i\alpha\sigma}$,
$\Delta_\alpha$ is the on-site energy on orbital $\alpha$ due to the crystal field splitting, and $t^{\alpha\beta}_{ij}$ are hopping parameters.
$U'=U-2J$ and $J=0.2U$ is the Hund's coupling.

The model in Eq. \ref{model} can in principle describe any two orbital systems as long as $\{t^{\alpha\beta}_{ij}\}$ are known. For a $d_{xz}$ and $d_{yz}$ two orbital system $H_0$ can be written as
\beq
H_0&=& \sum_{\vec{k},\sigma} \sum_{\alpha,\beta=1}^2\left[t^{\alpha\beta}_{\vec{k}} + \delta_{\alpha\beta}\delta_{ij}(\Delta_\alpha - \mu) \right]d^\dagger_{\vec{k}\alpha\sigma} 
d_{\vec{k}\beta\sigma}\nn\\
t^{11}_{\vec{k}} &=& - 2t_\parallel \cos k_x - 2t_\perp \cos k_y - 4t'\cos k_x \cos k_y,\nn\\
t^{22}_{\vec{k}} &=& - 2t_\perp \cos k_x - 2t_\parallel \cos k_y - 4t'\cos k_x \cos k_y,\nn\\
t^{21}_{\vec{k}} &=& t^{12}_{\vec{k}} = - 4t''\cos k_x \cos k_y,\nn\\
\label{t0}
\eeq
where $\alpha=1$ (2) represents the $d_{xz}$ ($d_{yz})$ orbital, and $t_{\parallel}$ ($t_{\perp}$) is the nearest neighbor hoppping via $\sigma$ ($\pi$) bonding respectively.

\begin{figure}
\includegraphics[width=3.4in]{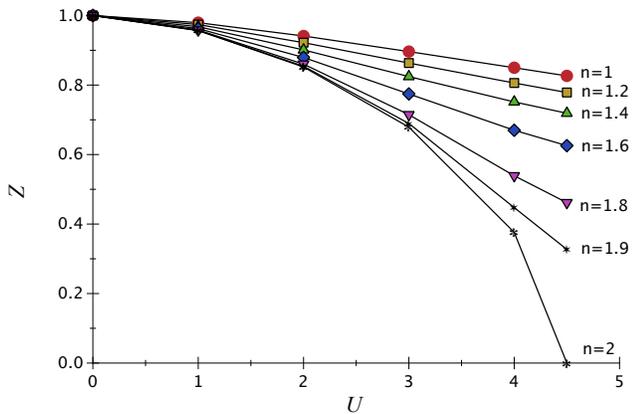}
\caption{\label{fig:nostrain} The quasiparticle weight in the case without the strain as a function of $U$ for different filling $n$. 
Because $d_{xz}$ and $d_{yz}$ remain degenerate, $Z_1=Z_2=Z$. The parameters in Eq.\ref{t0} are $t_\parallel =1$, $t_\perp = 0.1$, $t'$ = 0.5, and $t''=0.25$. The Mott transition ($Z=0$ at $n=2$) occurs
around $U_c=4.5$ for this choice of hopping parameters.}
\end{figure}

It is instructive to study the case with $C_4$ symmetry and $d = 0$. In this case, $d_{xz}$ and $d_{yz}$ orbitals are degenerate, and therefore we have $n_1 = n_2$ and $Z_1=Z_2=Z$. 
Clearly, at $n$ away from the half-filling, $Z$ remains close to 1 even for large $U$,
while at $n$ near the half-filling, $Z$ decreases much more significantly as $U$ increases. Fig. \ref{fig:nostrain} plots $Z$ as a function of $U$ for different values of filling $n$. 
The Mott transition occurs around $U/t_{\parallel} =4.5$ at exactly $n=2$.

\subsection{Experimental Details}
Epitaxial VO$_2$ films were grown on rutile (001), (100), and (110) oriented TiO$_2$ single crystal substrates by reactive MBE. Substrates were prepared by etching and annealing to have clean and well-defined step and terrace microstructured surfaces. Vanadium and distilled ozone were codeposited onto the substrate held at 250~$^\circ$C under a distilled ozone background pressure of 1.0$\times$10$^{-6}$~Torr. Following deposition of the desired 10~nm film thickness, the temperature of the sample was rapidly ramped to 350~$^\circ$C, then immediately cooled to below 100~$^\circ$C under the same background pressure of distilled ozone to achieve an improved film smoothness and a more abrupt MIT. For further details, the reader is referred to \textcite{paik2015}. 

\begin{figure}
\includegraphics[width=3in]{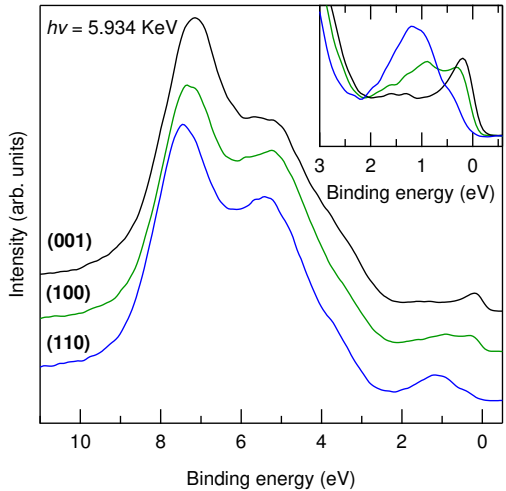}
\caption{\label{HAXPES} Full valence band HAXPES of VO$_2$(001), VO$_2$(100) and VO$_2$(110) collected above T$_{MIT}$.}
\end{figure}

The HAXPES measurements were performed using a photon energy of h$\nu$ = 5.934 keV and a pass energy of 200 eV, with a corresponding resolution better than 200 meV. The binding energy axes were referenced to both the Au 4f$_{7/2}$ and Fermi edge of a Au foil in electrical contact with the film. Figure \ref{HAXPES} displays the full valence band spectra of each film presented in the main text. The spectra are normalized to the O 2p states and offset for comparison.

\begin{figure}
\includegraphics[width=3.2in]{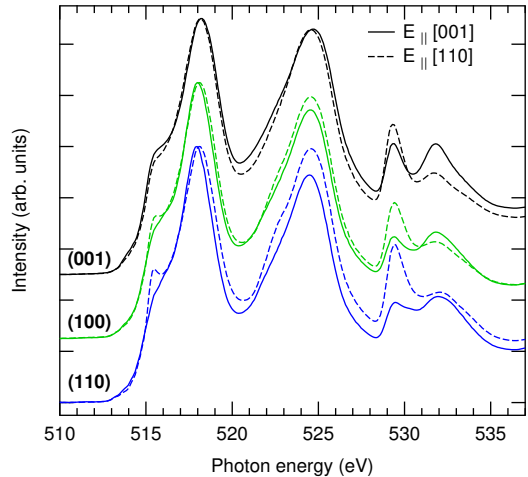}
\caption{\label{XAS} XAS of the V L-edge (513-528~eV) and O K-edge (528-536~eV) with polarization vector parallel (solid lines) or perpendicular (dashed lines) to the rutile c-axis of VO$_2$(001), VO$_2$(100) and VO$_2$(110) measured above T$_{MIT}$}
\end{figure}

The XAS was performed in total electron yield (TEY) mode by measuring the sample drain current and was normalized by the incoming beam current  with a linearly polarized beam. The photon energy axis was calibrated using the Ti L$_{2,3}$ and O K absorption edge features of a rutile TiO$_2$ single crystal. To achieve the desired alignment between the x-ray polarization and crystallographic direction, the films were rotated in both polar and azimuth. The spectra are normalized to the V L$_3$ peak to account for the intensity variations due to sample geometry.

The samples were heated with a resistive heating element while monitoring the O K-edge for signatures of the MIT, i.e. V--V dimers. Once the MIT was observed, samples were heated an additional $\sim$20~$^\circ$C above T$_{MIT}$ to avoid phase coexistance. Figure \ref{XAS} shows the polarization dependent XAS for VO$_2$(001), (100), and (110). The spectra includes both the V L (presented in the main text) and O K-edges. The O K-edge spectra presented here show no evidence V--V dimers, consistent with the rutile metallic phase.

\end{document}